\documentclass[aps,prb,floatfix,twocolumn,showpacs,tightenlines,groupedaddress,superscriptaddress,amsmath]{revtex4}

\pdfoutput=1

\usepackage{verbatim} 

\ifx\pdftexversion\undefined
  \usepackage[dvips]{color,graphicx}
\else
  \usepackage[pdftex]{color,graphicx}
\fi

 \usepackage{ifpdf}
    \ifpdf
 	 \usepackage[pdftex]{color,graphicx}
    \else
  \usepackage[dvips]{color,graphicx}
 \fi

\usepackage{amsmath}
\usepackage{amssymb}
\usepackage{subfigure}
\usepackage{array}
\usepackage{mathtools}
\usepackage{txfonts}
\DeclareMathAlphabet{\mathcal}{OMS}{cmsy}{m}{n}
 \newcommand{\ACcomment}[1]{}

\def	\bse{\begin{subequations}}
\def	\ese{\end{subequations}}

\newcommand{\dg}{\dagger}
\newcommand{\tc}{t_c}
\newcommand{\D}{\hat{d}}

\newcommand{\kd}{k_{\rm dot}}
\newcommand{\gd}{\gamma_{\rm dot}}
\newcommand{\st}{\hat{S}}
\newcommand{\z}{z_{\rm  m}}
\newcommand{\rt}{\hat{\rho}_{\rm rot}}

\newcommand{\de}{\delta}

\newcommand{\w}{\omega_m}

\def \dA{A_\delta}
\def \hzm{\hat{z}_{\rm m}}

\def \be{\begin{equation}}
\def \ee{\end{equation}}
\def \bew{\begin{widetext}\begin{equation}}
\def \eew{\end{equation}\end{widetext}}
\def \bmlett{\begin{mathletters}}
\def \emlett{\end{mathletters}}

\def \dg{\dagger}

\def \NN{{\mathcal N}}

\def \ra{\rightarrow}

\def \hF{\hat{F}}

\def \hH{\hat{H}}

\def \hn{\hat{n}}

\begin{document}

\title{Mechanically probing coherent tunnelling in a double quantum dot}

\author{J. Gardner }
\affiliation{Department of Physics, McGill University, Montr\'eal, Qu\'ebec, Canada H3A 2T8}

\author{S. D. Bennett}
\affiliation{Department of Physics, Harvard University, Cambridge, MA 02138, USA}

 \author{A. A. Clerk}
\affiliation{Department of Physics, McGill University, Montr\'eal, Qu\'ebec, Canada H3A 2T8}

%\keywords{Double quantum dot, artificial molecule, two-level system, back-action, electrostatic force microscopy, dissipation}
\date{Oct. 28, 2011}

\begin{abstract}
We study theoretically the interaction between the charge dynamics of a few-electron double quantum dot and a capacitively-coupled 
AFM cantilever, a setup realized in several recent experiments.  We demonstrate that the dot-induced frequency shift and damping of the cantilever can be used as a sensitive probe of coherent inter-dot tunnelling, and that these effects can be used to quantitatively extract both the magnitude 
of the coherent interdot tunneling and (in some cases) the value of the double-dot $T_1$ time.  We also show how the adiabatic modulation of the dot eigenstates by the cantilever motion leads to new effects compared to the single-dot case.
%This is possible despite the fact that the cantilever frequency is typically much smaller than relevant quantum dot frequency scales.

%Self-assembled quantum dots have been studied intensely because of their possible applications to quantum information processing. While such dots are difficult to characterize using direct electrical transport measurements, it has recently been shown both theoretically and experimentally that a capacitively coupled AFM cantilever can serve as a sensitive probe of dot charge dynamics and electronic level structure. This sensitivity is based on the fact that the dot, which is tunnel-coupled to electrons in a reservoir, acts as a dissipative bath for the cantilever. Here, we extend previous theoretical work to describe an AFM cantilever coupled to a double quantum dot. Unlike a single-dot, the double-dot system exhibits both incoherent tunneling to the leads and coherent tunneling between the dots. We find that the cantilever's motion is affected by both kinds of tunneling and can yield significant information even in regimes where the total double-dot charge does not fluctuate. Cantilever dynamics can also be used to learn about the strength of dephasing processes in the double-dot. After presenting the theoretical approach to this problem, we will discuss the results in the context of current experimental efforts using InAs dots. These effects should also be accessible in a variety of other quantum dot setups. 
\end{abstract}

\maketitle

% \tableofcontents

%-----------------------------------------------------------------------------------------------------------------------------------

Quantum dots have received significant attention both for their applications to quantum information and as laboratories for studies of fundamental physics.  Self-assembled, epitaxial quantum dots (QDs) offer advantages over lithographically defined dots in terms of size, confinement potential, and scalability \cite{Drexler94}.  Their small size however makes direct electrical characterization (e.g.~via transport) extremely difficult.  An alternate approach uses an oscillating atomic force microscope (AFM) tip which is only capacitively coupled to the QD charge \cite{McEuen02,zhu05, Stomp05,Cockins10,Dana05}.  The dot charge acts as a force on the cantilever, and hence its dynamics can alter the cantilever frequency and damping rate.  These effects provide detailed information on the dot, similar to that revealed by transport measurements or direct charge-sensing techniques (i.e.~using a nearby coupled electrometer).  It can even reveal subtle effects involving the interplay between orbital degeneracy and Coulomb blockade physics that are difficult to obtain by other means \cite{Cockins10,Bennett10}. 
%the dot charge and the tip-sample force.  Changes in the resonant frequency and damping of the oscillator reveal corresponding susceptibilities in the charge.  Recent theoretical and experimental work has emphasized the value of such an approach not only for detecting single-electron charge transport, but also as a sensitive probe of QDs' atom-like properties, such as orbital degeneracy \cite{Cockins10,Bennett10}.  The limitations of conductance-based measurements are compounded for more complex nanostructures such as double quantum dots (DQDs), where new degrees of freedom require more intricate characterization techniques \cite{Amaha08}.  Electromechanical measurements, by contrast, function with undiminished sensitivity in complex, multi-dot environments \cite{Cockins10}.  Indeed, it may be in precisely these situations that electromechanical methods reveal their true potential.

In this work, we focus theoretically on new effects that arise when a low-frequency cantilever is coupled to a double quantum dot (DQD)  (i.e.~two QDs which are coupled capacitively and via coherent tunnelling \cite{vanderWiel02,Wang09}).  Unlike the single-dot case, the cantilever is now sensitive to variations in the distribution of charge between the two dots.  We find that
this sensitivity leads to new mechanisms for a dot-induced cantilever damping and frequency shift.  These effects are not solely the consequence of incoherent tunnelling to a reservoir (as is the case for a single dot), but instead depend sensitively on the strength of coherent tunnelling between the quantum dots.  Our results are derived using a linear-response,  quantum master-equation calculation; this extends the semi-classical Fokker-Planck treatments used so often in quantum electromechanics \cite{Blencowe04,Bruder04,Blanter04, Cockins10,Pistolesi07,Rodrigues05,Doiron06} to now include coherent interdot tunneling.

The most prominent new effects emerge in the vicinity of the so-called charge transfer line, where two DQD charge configurations having the same total charge are almost degenerate.  
%This near-degeneracy is lifted by coherent interdot tunneling.  
In this regime, we find a new mechanism for DQD-induced cantilever damping that is enhanced by the relatively long time scale for charge relaxation.  We show that this effect can be used to directly measure the $T_1$ time of the DQD.    We also find a new mechanism for a dot-induced cantilever frequency shift near the charge transfer line.  In this regime, the presence of coherent tunneling implies that the DQD energy eigenstates are superpositions of charge eigenstates.  The oscillator motion can adiabatically modulate the corresponding wavefunctions, which gives rise to the new frequency-shift mechanism.  Not surprisingly, as this effect is a direct consequence of having superpositions of charge states, it can be used to probe the strength of coherent interdot tunneling.

%This paper focuses on the physics behind non-contact electromechanical measurements of charge dynamics in double dot systems.  Coupled to a single dot, the cantilever is sensitive only to changes in the total dot charge. In the double dot case, the oscillator can also detect fluctuations in the charge distribution---sometimes in dramatic ways.  Even for fixed total charge, it may be possible to observe anomalously large damping due to slow, incoherent changes in the single-electon probability distribtion.  Moreover, the resonant frequency of the cantilever could contain evidence of coherent mechanical coupling to the eigenstate wavefunctions of the DQD.  The surprising character of these results is rooted in the physics of electron tunneling between the coupled quantum dots, which depends on both incoherent and coherent processes.  Using simple physical arguments to support more comprehensive analytic results, we show how non-contact AFM measurements could provide a uniquely powerful tool for probing electron transport in double quantum dots.

%\section{Model}
{\it Model--}  Motivated by experiments \cite{Cockins10,Bennett10}, we consider a setup where a self-assembled DQD sitting on a surface is capacitively coupled to an oscillating metallized cantilever; the dots are also tunnel-coupled to a two-dimensional electron gas (2DEG) sitting below the surface.  
The DQD is described by a standard Coulomb blockade Hamiltonian, plus a term describing coherent interdot tunneling (strength $t_c = |t_c| $).  
% As we focus on the experimentally-relevant regime of a small-cantilever frequency, we begin by treating the cantilever position $z_m$ as an external (classical) parameter. 
We will be interested in the few-electron regime where each dot has at most one electron, and thus retain only a single orbital in each dot; for simplicity, we also treat the case of spinless electrons, as including spin does not significantly change our results.  

For a fixed cantilever position, we have:
\begin{equation}
	\hat{H}_{\rm DQD} =  
		\hat{H}_c+t_c \D_L^\dg \D_R + t_c \D_R^\dg \D_L +\hH_{\rm res} \label{eq:Hdqd}
\end{equation}
where $\D^\dg_{\rm \alpha}$ is the electron addition operator for dot $\alpha=L,R$.  The 
charging Hamiltonian $\hat{H_c}$ takes the standard form \cite{vanderWiel02}:
\begin{equation}
	\hat{H}_c  =
		\sum_{\alpha=L,R} E_{\rm C\alpha }\left(\hat{n}_\alpha-\mathcal{N}_\alpha \right)^2
	+ E_{\rm Cm}\left(\hat{n}_L-		\mathcal{N}_L\right)\left(\hat{n}_R-\mathcal{N}_R\right). \label{eq:Hc}
\end{equation}
$E_{\rm C\alpha}$ is the charging energy of dot $\alpha$, $E_{\rm Cm}$ describes their capacitive coupling, and $\hat{n}_{\rm \alpha}=\D_{\rm \alpha}^\dg \D_{\rm \alpha}$ is the dot $\alpha$ electron number operator.  We focus exclusively on the Coulomb blockade regime where $E_{\rm C \alpha}>>k_BT$.
Finally, $\hH_{\rm res}$ describes the 2DEG as a free electron gas, and dot-2DEG tunneling.  We take the 2DEG to be in equilibrium at temperature $T$, and assume for simplicity that the 2DEG-dot tunnel matrix element is the same for both dots.  We use $\Gamma$ to denote the maximum Golden rule tunnel rate from a given dot to the 2DEG.

The only parameters in $\hat{H}_{\rm DQD}$ depending on the cantilever position $\vec{r}_{\rm tip}$ are the dimensionless gate voltages 
$\NN_{\rm \alpha} = -V_B C_{ \rm tip,\alpha} ( | \vec{r}_{\rm tip} - \vec{r}_\alpha|) /e$,
 where $C_{ {\rm tip} , \alpha} $ is the cantilever-dot capacitance, $\vec{r}_\alpha$ denotes the position of dot $\alpha$, and $V_B$ is the voltage applied between the cantilever and the 2DEG.  As demonstrated in Refs.~\onlinecite{Cockins10,Stomp05,Dana05}, by varying the tip position at a fixed height above the DQD sample plane, one effectively varies 
 $\NN_L, \NN_R$ and thus maps out the well-known DQD ``stability diagram" \cite{vanderWiel02} (i.e~realizes different ground-state charge configurations).
 One can thus effectively view $\NN_L, \NN_R$ as independent parameters, similar to conventional gated devices.
    % Though the exact dependence of gate voltage on tip position will vary with each dot, the isomorphism between tip position and gate voltage permits us to think of the $\NN_\alpha$ as independent.  We will therefore often use the stability diagram format to simplify discussions of results.  

We now allow the cantilever height $\z$ to oscillate, letting $\z=0$ denote its equilibrium position.  The co-ordinate $\z$ is well-described as a harmonic oscillator having a frequency $\w$ and mass $m$.  The coupling between the oscillations and the DQD electrons arises solely through the dependence of the tip-sample capacitance (and hence $\NN_{\alpha}$) on $\z$.  For the small oscillations of interest, we can linearize this dependence. Eq.~(\ref{eq:Hc}) then yields the DQD-cantilever interaction Hamiltonian:
\begin{eqnarray}
	\hH_{\rm int} & = & 
		-\hzm \displaystyle\sum_{\rm \alpha=L,R} A_{\rm \alpha}\hat{n}_{\rm \alpha} 
%		= -\z \left[ \bar{A} (\hn_L + \hn_R) + (dA) (\hn_L - \hn_R ) \right]
		\equiv  -\hzm	\hat{F}, 
		\label{eq:int}	\\
	A_{\rm \alpha} & = & 
		2 E_{\rm C\alpha}\frac{\partial\mathcal{N}_{\rm \alpha}}{\partial \z}
		+E_{\rm Cm}\frac{\partial\mathcal{N}_{\rm \bar{\alpha}}}{\partial \z} 
		=	-\frac{\partial E_{+ \alpha}}{\partial \z}, \label{eq:int1}
\end{eqnarray}
where $\alpha=L,R$ and $\bar{\alpha}=R,L$ are complementary indices.  In the last equality, we have introduced the electron addition energies 
$E_{\alpha +} = 
E_{\rm C \alpha} \bigl ( 1-2 \mathcal{N}_{\alpha} \bigr) -  E_{\rm C m} \mathcal{N}_{\bar{\alpha}}$ 
associated with adding a single electron to dot $\alpha$ to an initially empty DQD state (i.e.~zero electrons in either dot).  
%As in general $A_l \neq A_r$,  the force $\hF$ exerted by the DQD on the cantilever depends on both the total charge in the DQD as well as $\hn_L - \hn_R$.

%Starting from the state with no extra electrons ($|00\rangle$), The last equality in Eq.~(\ref{eq:int1}) establishes the direct effect of $\z$ on the DQD addition energies.  Note that there are other order-$\z$ terms in the system energy that we ignore because they are independent of the charge fluctuations we wish to investigate \cite{Stomp05}. 
%\begin{figure*}[t!]
%	\includegraphics[angle=0,bb=0 0 500 230]{./img/stability_both.png}
%	\caption{Stabilility diagrams depict the thermal charge configuration at $T = 10 K$ as a function of tip position (a) and gate voltage (b).  The color scale indicates the average total charge, which varies as a thermally broadened step function between integer values.  The regions are labeled by ordered pairs $(\langle \hat{n}_l\rangle,\langle \hat{n}_r\rangle)$ corresponding to the ground state configuration.  The two one-electron regions are divided with red dotted \textit{charge transfer lines}.  In (a), the locations of the dots are marked by 'x' symbols.  These plots are generated using the following parameters: $t_c = 0.5\: meV$, $E_{cl} = E_{cr} =31\: meV$, $E_{m} = 20\: meV$, and $V_B = -15 \:V$.}
%	\label{figure:stab}
%\end{figure*}

We are most interested in the regime where the total DQD charge is fixed at one,
and where the electrostatic energy detuning  $\de  =  ( E_{+L}-E_{+R}) /2  $ of the two relevant charge states
$|10\rangle$ (electron on left) and $| 01\rangle$ (electron on right) is small.  In this regime, we can safely neglect charge states where the total DQD charge is larger than 2.  Further, we will focus on DQDs where the coherent tunneling $t_c$ is much larger than both the DQD-2DEG tunneling rate $\Gamma$ and the mechanical frequency $\w$; this condition is readily achieved in self-assembled QDs (cf.~Ref.~\onlinecite{Amaha08}).  It is thus useful to work in the basis of adiabatic eigenstates:  the eigenstates of   
$\hH_{\rm DQD}$ determined by the instantaneous value of $\langle \hat{z}_m \rangle$.  Two of the four eigenstates are simply charge
eigenstates: $|1[\z]\rangle = | 11 \rangle$, $|4[\z]\rangle = | 00 \rangle$.  For the remaining eigenstates, note that
Eq.~(\ref{eq:int}) implies that the detuning $\delta$ will vary linearly with $\z$.
We thus define:
\begin{equation}
	\theta[\z] = 
%	\arctan \left[   \frac{2 t_c }{ E_{+L}[\z]-E_{+ R}[\z] } \right] \equiv 
		\arctan \left[ \frac{ t_c }{\de[\z]} \right] 
	=  		\arctan \left[ \frac{  t_c }{\de - \z(A_{L}-A_{R})/2} \right] . \label{eq:theta}
%	\tan\bigl(\theta[\z]\bigr)=\frac{2|t_c|}{E_{+l}[\z]-E_{+ r}[\z]}\equiv \frac{|t_c|}{\de} .
\end{equation}
The remaining adiabatic eigenstates are thus:
\bse
\label{eqs:AdiabaticEigenstates}
\begin{eqnarray}
	| 2[\z] \rangle & \equiv & | g[\z]\rangle=-\sin\left(\theta/2\right)\: |10 \rangle + \cos\left(\theta/2\right)\: |01 \rangle \label{eq:basis1} \\
	| 3[\z] \rangle & \equiv &  | e[\z]\rangle=\cos\left(\theta/2 \right) \: |10 \rangle +\sin\left(\theta/2\right) \: |01 \rangle,  \label{eq:basis2}
\end{eqnarray}
\ese
with corresponding adiabatic eigenenergies 
\begin{eqnarray}
	E_{\rm  g,e}[\z] & = & \left(\frac{E_{+L}+E_{+R} - (A_L + A_R) \z }{2} \right) \!\mp
		 \sqrt{ \left[ \de[\z] \right]^2 + t_c^2 } 
		\nonumber \\
		& \equiv & 	
	\bar{\epsilon}[\z]  \!\mp \Delta[\z].
	\label{eq:DeltaDefn}
\end{eqnarray}
For the low temperatures we focus on, the DQD will primarily occupy the states $|2[\z] \rangle $ and $|3[\z] \rangle $, and thus will approximate the physics of a two-level system.
%where $\Delta[\z]$ represents the splitting of the adiabatic eigenstates and $\bar{\epsilon}[\z]$ is their average value.

% ------------------------------------------------------------------------------------------------------------
% ------------------------------------------------------------------------------------------------------------
% ------------------------------------------------------------------------------------------------------------

{\it Calculation-- }
As a result of the coupling in Eq.~(\ref{eq:int}), the average force from the dot $\langle \hF \rangle$ will respond with a delay to the motion of the oscillator resulting in both a spring-constant shift $\kd$ and extra damping $\gd$; for the weak couplings of interest, this is fully described within linear response \cite{BennettClerk05}.  
To quantitatively describe these effects in the regime 
$\hbar \w \ll k_B T$, we derive a Lindblad master equation describing the DQD and cantilever.  For a  single-QD system, this approach yields a classical master equation with incoherent tunnelling rates set by the instantaneous oscillator position \cite{Cockins10,Blencowe04,Bruder04,Blanter04, Doiron06, Rodrigues05,Pistolesi07}.  To extend this approach to include coherent tunneling, note that since we will calculate $\kd$ and $\gd$ within linear response, we can without loss of generality replace the cantilever position $\hat{z}_m$ by its average value:  
$\hat{z}_m \ra \z(t)=z_{\rm  0}\cos(\w t)$.  Understanding how the DQD responds to this time-dependent classical field will then yield $\kd$ and $\gd$.
Defining $\hat{U}[x]$ via $\hat{U}[x] |j[0] \rangle =   |j[x] \rangle$, 
we define  $\rt(t)= \hat{U}[\z(t)]^\dagger \hat{\rho}(t)\:    \hat{U}[\z(t)]$, 
where $\hat{\rho}(t)$ is the Schr\"odinger-picture reduced density matrix of the DQD.  $\rt$ is simply the DQD density matrix in the adiabatic basis.  
Using the experimentally-relevant conditions $|t_c|>>\hbar \Gamma $ and $(\bar{A}_{L,R}\: z_{\rm  0} \hbar \w) <<(k_BT)^2 $, and making
Born-Markov and secular approximations, we obtain:
% the master equation for $\rt$ takes the form :
\begin{equation}
	\frac{\partial\rt}{\partial t}=\frac{1}{i \hbar}[\hat{H}_{\rm eff},\rt]+\displaystyle\sum_{ j,k=1}^4 \Gamma_{\!jk}\mathcal{D}[\st_{\!jk}]\rt. \label{eq:lindblad}
	\end{equation}
The first term on the RHS describes coherent evolution. Using Pauli matrices to describe the states 
$|2\rangle \equiv |2[0] \rangle$, $|3\rangle \equiv |3[0] \rangle $ as a two-level system in the natural way, e.g.
~$\hat{\sigma}_z = |3 \rangle \langle 3| - |2 \rangle \langle 2| $, $\hat{\sigma}_x = |3 \rangle \langle 2| + |2 \rangle \langle 3| $, 
$\hat{\sigma}_y = i( |2 \rangle \langle 3| - |3 \rangle \langle 2|)$, we find:
\begin{eqnarray}
	\hat{H}_{\rm eff}  & = & 
		\bar{\epsilon} \:\hat{n}_{\rm tot} + E_m |4\rangle\langle 4|+  \Delta \:\hat{\sigma}_z -
\frac{\hbar}{2}\frac{\partial \z}{\partial t}\frac{\partial \theta}{\partial \z}\:\hat{\sigma}_y, \label{eq:eff}
%	B = = z_{\rm   0}\w\frac{\da|t_c|}{\Delta^2}. \label{eq:fict}
\end{eqnarray}
The last term here describes an effective state precession; its origin is the rotation of the adiabatic eigenstates brought on by $\z(t)$. Note that similar adiabatic approaches have been used to study periodically-modulated, dissipative two-level systems, with the dissipation being treated phenomenologically \cite{Jackle76,Golding73,Hunklinger76,Stockburger95,parshin93}, or as a bosonic bath \cite{Grifoni98}.  In contrast, our system is effectively a four state system, and the dominant dissipation due to 2DEG tunneling is treated microscopically. 

The remaining terms on the RHS of Eq.~(\ref{eq:lindblad}) have the standard form of Lindblad dissipation.  Letting $\st_{\!jk} = |j\rangle\langle k|$, 
the superoperators $\mathcal{D}[\st_{ \!jk}]$ are defined as:
\begin{equation}
	\mathcal{D}[\st_{ \!jk}]\rt=\st_{\! jk} \rt \st_{\! jk}^{\dagger}-\frac{1}{2}\left( \st_{\! jk}^{\dagger}\st_{\! jk}\:\rt + \rt \:\st_{\!jk}^{\dagger}\st_{\! jk}\right), 	\label{eq:jumpt}
\end{equation}
In the case where the total charge in states $|j\rangle$ and $|k\rangle$ differ by $1$, $\Gamma_{\!jk}$ is simply a Fermi Golden rule rate for DQD-2DEG tunneling determined by the {\it instantaneous} eigenstate energies $E_{j}[\z(t)]$, $E_{k}[\z(t)]$.  
In contrast, the rates $\Gamma_{23}, \Gamma_{32}$ describe the intrinsic relaxation of the DQD (e.g.~due to electron-phonon interactions), with $1/T_{1,\rm int} \equiv \Gamma_{23} + \Gamma_{32}$.
We will treat such processes phenomenologically by taking $1/T_{1,\rm int}$ to be a parameter.  We also assume that the bath responsible for the intrinsic relaxation has the same temperature as the 2DEG; as such, the $\z=0$ stationary solution of Eq.~(\ref{eq:lindblad}) is simply a thermal occupation of the states $| j \rangle$.
To find the dot-induced damping and spring-constant shift of the cantilever, we use Eq.~(\ref{eq:lindblad}) to find the first-order-in-$\z$ correction to $\rt$, and use the corresponding change in 
$\langle \hF(t) \rangle$ to get $\gd$ and $\kd$ in the standard manner (see, e.g., Ref.~\onlinecite{BennettClerk05}).

% I don't think the paragraph below is needed; unecessary confusing.
%Changes in the total DQD charge are effected by electrons tunneling to and from the 2DEG.  For the simplest case where lead tunneling is symmetric between the dots and tunneling electrons do not interfere, the lead coupling can be defined by $\Gamma = \Gamma_{1\beta}+\Gamma_{\beta 1} = \Gamma_{2\beta}+\Gamma_{\beta 2}$, where $\beta=2,3$.  Note that for our parameters (which correspond to a slow oscillator), the incoherent lead tunneling is described by simple Lindblad ``jump" operators between adiabatic eigenstates; also, as the 2DEG is in thermal equilibrium, the ratio of these rates obeys detailed balance.
%For $z=0$, the fact that the 2DEG is in equilibrium implies that the stationary solution of Eq.~(\ref{eq:lindblad}) describes a thermal state.  

%:
%\begin{eqnarray}
%	\kd & = &
%		 \frac{\w}{\pi z_0}\displaystyle\int_0^{2\pi/\w} dt \delta\langle \hat{F}(t) \rangle\: \z(t) 
%		 	\label{eq:shift1} \\
%	\gd & = &
%		 \frac{-1}{m\pi z_0}\displaystyle\int_0^{2\pi/\w} dt \delta\langle \hat{F}(t) \rangle \:\dot{z}_{\rm  m}(t). \label{eq:damp}
%\end{eqnarray}

% ------------------------------------------------------------------------------------------------------------
% ------------------------------------------------------------------------------------------------------------
% ------------------------------------------------------------------------------------------------------------

{\it Basic mechanisms- }In the low-frequency limit, the linear-response results take the form:
\begin{eqnarray}
	m \gd = \tau \frac{ \partial \langle \hF \rangle }{\partial \z}	,\:\:
	\kd = - \frac{ \partial \langle \hF \rangle }{\partial \z}	\label{eq:heuristic}
\end{eqnarray}
where $\tau$ is an effective response time \cite{BennettClerk05}. 
\ACcomment{Note that $\gd$ and $\kd$ necessarily have opposite signs.}
In the single dot case, the static susceptibility $\partial \langle \hF \rangle/\partial \z$ is just proportional to the charge susceptibility $\partial \langle \hn_{\rm tot} \rangle / \partial \NN$.  $\gd$ and $\kd$ are thus only significant when the QD is tuned to a point where its total charge can fluctuate via 2DEG-QD tunneling; correspondingly,  $\tau \sim 1 / \Gamma$\cite{Bennett10}.  In contrast, these fluctuations of total charge are exponentially suppressed in a DQD near the charge transfer line.  As we now show, $\gd$ and $\kd$ are instead determined by the response and dynamics of the DQD charge distribution.

In the vicinity of the charge transfer line, the DQD-induced force operator $\hF$ takes the form ($\dA = (A_L-A_R)/ 2$): 
\begin{eqnarray}
	\hF - \frac{ A_L + A_R}{2} \simeq
%		\bar{A} + 
		\dA \left( \hn_L - \hn_R \right) 
		=  
%		\bar{A} + 
		\dA \left( \cos \theta \:\hat{\sigma}_z - \sin \theta \:\hat{\sigma}_x \right) \label{eq:hf}
\end{eqnarray} 
It follows that both $\gd$ and $\kd$ will have contributions from three distinct mechanisms, corresponding to the susceptibilities $\partial_{\z} \langle \hat{\sigma}_x \rangle$, $\partial_{\z} \langle \hat{\sigma}_z \rangle$ and $\partial_{\z} \theta $.  The first of these involves the oscillator motion modifying the coherence between the $|e\rangle$ and $|g\rangle$ DQD eigenstates.  This corresponds to the well-known resonant-damping mechanism of acoustic vibrations by a two-level system \cite{Remus09, Hunklinger76,Golding73}, and is strongly suppressed in our system as $\hbar \w \ll t_c$; we thus do not discuss it further.  The remaining two mechanisms are important for our system, and we discuss their effects in turn.  
% 	\begin{figure}[h!]
% 	\includegraphics[width=0.5\textwidth]{./img/shiftplace.pdf}
% 	\caption{The shift in the resonant frequency of an oscillator due to coupling to a DQD in position space (inset) and voltage space.  Charging lines dominate both images, but in the single electron regime near the charge transfer line it is the adiabatic contribution that determines the frequency shift.  These plots are generated using parameters XXX.}
% 	\label{figure:shift}
% 	\end{figure}

% ------------------------------------------------------------------------------------------------------------
% ------------------------------------------------------------------------------------------------------------
% ------------------------------------------------------------------------------------------------------------

\begin{figure}[t!]
	\includegraphics[width=8.25cm]{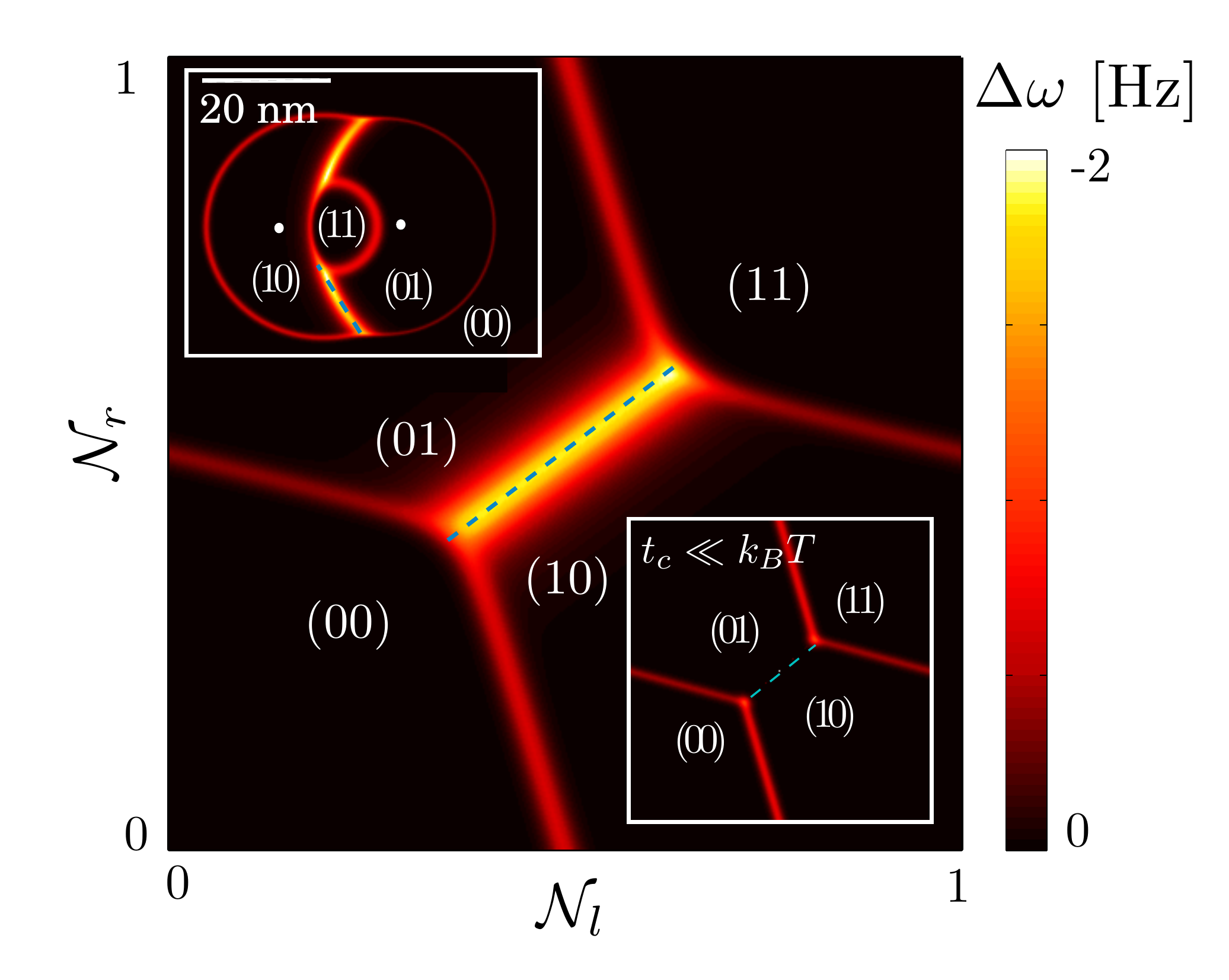} 
	\caption{Calculated DQD-induced frequency shift $\Delta \omega = \kd / (2 m \w)$, using parameter values similar to the experiment
	of Ref.~\onlinecite{Cockins10}.  
	Main plot:  frequency shift versus dimensionless gate voltage for $\w$~=~160~kHz, $k_0$~=~7~N/m, 
	$\Gamma$~=~10~kHz, $T$~=~4.2~K, $t_c$~=~1~meV, $E_{CL}$~=~20~meV, $E_{CR}$~=~25~meV, $E_{Cm}$~=~12~meV.
%	Features are seen at charge addition lines, boundaries between states having different total charge.  
	Most interesting is the feature
	along the charge transfer line (indicated with a dashed line), 
	which results from the adiabatic modulation of the DQD eigenstates by the cantilever.  The width of this feature
	is a measure of $t_c$.  For simplicity, we have used fixed couplings $A_L \approx 8$~meV/nm, $A_R \approx 6$~meV/nm.  Lower inset: Same as main plot, but now $t_c = 0.001 \:{ \rm{meV} } \ll k_BT$.  The adiabatic feature is absent.  Upper inset: Simulated AFM data of frequency-shift versus lateral tip position, for a tip height of $20$~nm and a bias voltage $V_B = -7.1$~V; white dots indicate the centres of the two dots.  The parameters and coherent tunneling are the same as the main plot, but the couplings $A_L, A_R$ now vary with tip position.  See Appendix A for more details.  
	 }
	\label{figure:shift}
\vspace{-0.5cm}
\end{figure}

%%%%%%%%%%%%%%%%%%%%%%%%%%%%%%%%%%%%%%%%%%%%%
%%%%%%%%%%%%%%%%%%%%%%%%%%%%%%%%%%%%%%%%%%%%%
%%%%%%%%%%%%%%%%%%%%%%%%%%%%%%%%%%%%%%%%%%%%%
{\it Adiabatic frequency shift-- } Eq.~(\ref{eq:hf}) implies that as $\hF$ has an explicit dependence on $\theta$, the intrinsic $\z$-dependence of $\theta$ (cf.~Eq.~(\ref{eq:theta})) will cause a modulation of $\hF$.  Physically, this corresponds to the adiabatic modulation of the DQD eigenstates by the cantilever oscillation (via the cantilever's modulation of the electrostatic detuning $\delta$).  The corresponding oscillation in $\langle \hn_L - \hn_R \rangle$ causes a force oscillation which is in phase with $\z(t)$; it thus contributes to the DQD-induced spring-constant shift $\kd$.  One finds simply:  
\begin{equation}
	k_{\rm dot, ad} =  - \dA \langle \hat{\sigma}_z \rangle \frac{\partial \cos\theta}{\partial \z} = \frac{-\dA^2\sin^2\!\theta\:\tanh(\Delta/k_BT)}{\Delta}.
	\label{eq:kdotad}
\end{equation}
where the RHS should be evaluated at $\z = 0$.  Note that as we focus on a small cantilever frequency ($\hbar\w \ll t_c^2/(z_0\: \dA)$), non-adiabatic Landau-Zener transitions will have negligible probability and can be safely neglected.  Such non-adiabatic transitions were recently studied in a two-mode optomechanical system \cite{Heinrich10}; unlike our work, the focus was on the regime where $\omega_m$ was much larger than the effective tunneling $t_c$.  We stress that $k_{\rm dot, ad}$ is a direct consequence of having coherent interdot tunneling, and vanishes in the limit $\tc \ra 0$.  It is most pronounced at low temperatures ($k_B T < t_c$), where it gives rise to a feature near the charge transfer line whose width (in $\delta$) is $\sim t_c$.  Further, at such low temperatures this effect dominates all other contributions to $\kd$ near the charge transfer line.  It thus provides a direct means for both detecting the presence of coherent interdot tunneling, and for measuring its magnitude.  

Shown in Figure~\ref{figure:shift} is a full calculation of the DQD-induced frequency shift $\Delta \omega = \kd /(2 m\w)$ obtained using Eq.~(\ref{eq:lindblad}) and linear response, keeping all contributions.  We have used experimentally-relevant DQD and cantilever parameters; see caption for details.  The lower inset shows results for a small value of $t_c$; similar to a single QD system, the only appreciable frequency shift occurs at charge addition lines where lead tunneling is strong and the total dot charge can fluctuate.  In contrast, for a larger value of $t_c$, one obtains a large frequency shift along the charge transfer line, in agreement with Eq.~(\ref{eq:kdotad}).  Again, seeing this effect provides a direct probe of coherent interdot tunneling.  Note that there exists a somewhat similar adiabatic contribution to the TLS - acoustic wave interaction in glasses \cite{Stockburger95}, but that it is neglected in the standard early treatments \cite{Jackle76,Golding73,Hunklinger76}.

\begin{figure}[t!]
	\includegraphics[width=8.25cm]{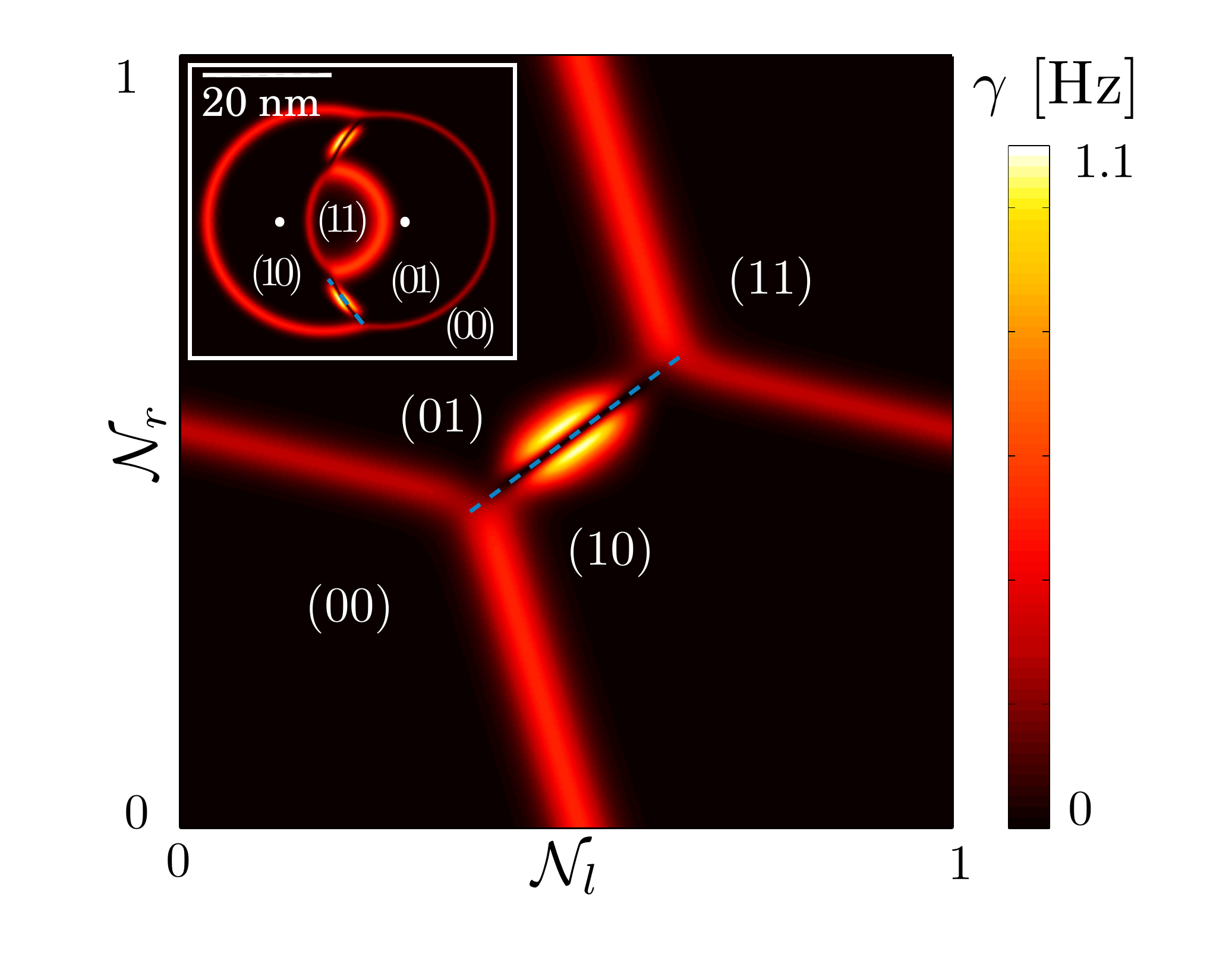} 
	\caption{
	Calculated DQD-induced damping $\gd$.  DQD parameters are identical to Figure~\ref{figure:shift}, except for $t_c = 0.3$~meV, $T = 8.4$~K, and $\Gamma = 10$~MHz.  We have also included an intrinsic relaxation mechanism ($T_{\rm 1,int}^{-1}\sim 100\:kHz$) which imposes a lower bound on the TLS relaxation rate.  The AFM tip parameters are $\w = 75$~kHz and $k_0=3$~N/m.  The main plot shows the damping as a function of the dimensionless gate voltages, while the inset shows a simulated AFM damping-versus-position image.  The damping features near the charge transfer line correspond to the TLS damping mechanism discussed in the text.  
	 }
	\label{figure:damping}
	\vspace{-0.5 cm}
\end{figure}

% ------------------------------------------------------------------------------------------------------------
% ------------------------------------------------------------------------------------------------------------
% ------------------------------------------------------------------------------------------------------------

{\it Effective TLS damping-- }Eq.~(\ref{eq:hf}) indicates a second mechanism which contributes to $\kd$ and $\gd$ near the charge transfer line: the cantilever's modulation of $\langle \sigma_z \rangle$, the population asymmetry of the two low-energy DQD eigenstates.  This corresponds directly to the well-known mechanism of non-resonant damping by a two-level system (TLS), studied in the context of acoustic damping in glasses \cite{Hunklinger76, Jackle76, Golding73, Cleland03, Grifoni98,Stockburger95,parshin93}.  On a heuristic level, the cantilever oscillations cause the DQD splitting $\Delta$ to oscillate (cf.~Eq.~(\ref{eq:DeltaDefn})), which in turn causes the occupancy of the states $|2 \rangle$, $|3 \rangle$ to oscillate.  The corresponding oscillations in $\langle \hat{\sigma}_z\rangle$ 
(and hence $\langle \hF \rangle$) are phase shifted with respect to $\z(t)$ due to the finite DQD $T_1$ time; the mechanism thus contributes both to 
%the damping 
$\gd$ and 
%spring-constant shift 
$\kd$.
This mechanism for damping relies on the DQD being coupled to a bath (e.g.~the reservoir electrons) 
which allows its populations to re-equilibrate in response to changes in the splitting energy $\Delta$; hence, the energy dissipated is ultimately transferred to this bath.
Note also that this mechanism is suppressed at low temperatures $k_B T \ll \Delta$, as in this case the DQD is always in its ground state.

We find that the DQD-induced damping due to this process is given by:
\begin{equation}
	 m \gd = \left(\frac{T_1}{1+\w^2T_1^2}\right) \frac{\dA^2 \cos^2\!\theta}{k_B T \cosh^2(\Delta / k_B T)} \label{eq:tlsd},
\end{equation}
in agreement with previous works  \cite{Hunklinger76, Jackle76, Golding73}.  
Unlike previous works, in this system one knows the precise microscopic nature of the TLS (i.e.~an electron in the DQD), 
and also knows at least some of the processes
contributing to its relaxation time $T_1$.  For our model of spinless electrons, we find:
\begin{equation}
	T_1^{-1} = \Gamma\cosh(\beta\Delta)\Bigl( e^{-\beta|\bar{\epsilon}|} + e^{\beta(|\bar{\epsilon}|-E_m)}\Bigr)+T_{1,{\rm int}}^{-1}, \label{eq:t1}
\end{equation}
where $\beta=1/(k_BT)$.  The first term in the relaxation rate corresponds to relaxation processes involving 2DEG-DQD tunneling and an intermediate state where the total DQD charge is either $2$ or $0$; note that this term depends explicitly on both $\Delta$ and $\bar{\epsilon}$, and will thus vary as one moves (in gate voltage space) along and away from the charge transfer line.  The form of this term corresponds to the simple case of spinless electrons and equal dot-2DEG tunnel couplings; the more general form is given in Appendix B.
The second term in Eq.~(\ref{eq:t1}) describes intrinsic relaxation processes in the DQD (e.g.~due to a coupling to phonons). 
Note that for this mechanism (i.e.~the $\hat{\sigma}_z$ contribution to $\hF$), $T_1$ plays the role of the response time $\tau$ in Eq.~(\ref{eq:heuristic}).

One might na\"ively think that significant dot-induced damping could only occur near charge addition lines where DQD-2DEG tunneling is strong.  However, 
for a low frequency cantilever, we see that near the charge transfer line, $\gd$ scales as $T_1$; in contrast, the more conventional $\gd$ mechanism near a charge addition line scales as $1 / \Gamma$.  Thus, if $T_1 \Gamma > 1$, this ``TLS damping" mechanism can be equal to or even greater in magnitude than the more conventional damping peaks found near charge addition lines.  This behaviour is shown in Figure~\ref{figure:damping}, where we use our full calculation to plot the DQD contribution to the cantilever damping, $\gd$.  It is interesting to note the presence of coherent tunneling causes the effect to vanish at $\delta = 0$, as $\Delta$ has no linear dependence on $\z$ here.  One can thus use the suppression of this damping effect on the charge transfer line as a direct probe of coherent interdot tunneling.

{\it Measuring $T_1$-- }In the simple case of a low frequency cantilever and a single mechanism contributing to both $\kd$ and $\gd$, Eq.~(\ref{eq:heuristic}) suggests that one can simply measure the relevant response time $\tau$ by taking the ratio of the two effects, without having to precisely know the strength of the dot-cantilever coupling.  A similar approach can be used to extract the DQD $T_1$ time near the charge transfer line, though more care is needed, as there are two mechanisms contributing to $\kd$.  First, note that the $\gd$ as given by Eq.~(\ref{eq:tlsd}) is only appreciable for $k_B T \gtrsim \Delta$.  For $k_B T \gg \Delta$, one finds that the two spring constant mechanisms combine in a particularly simple manner, and that the damping versus spring constant shift ratio takes the simple form:
%
%
%
%We now consider the feasibility of combining the damping and the spring constant shift to extract a measurement of the $T_1$ time for an isolated DQD.  We might first try calculating $T_1$ from Eq.~(\ref{eq:tlsd}), but this requires knowledge of the coupling asymmetry, $dA$---a difficult quantity to measure.  To circumvent this, consider instead the ratio $m\gd/\kd$, which is independent of $dA$.  In the simple, single dot case, this ratio would yield $\tau$ directly [cf. Eq.~(\ref{eq:heuristic})].  The double dot system is more complex because of the various susceptibilities, each of which corresponds to a different response time.  Nevertheless, our understanding of $\gd$ and $\kd$ is sufficient to obtain $T_1$ in the one-electron regime with no knowledge of $dA$.  A particularly simple case occurs when $\Delta << k_BT$.  The total frequency shift then becomes:
%\begin{equation}
% \kd \approx  -\beta(dA)^2.
%\end{equation}
%This leads to a very basic relationship between the measured quantities and $T_1$:
\begin{equation}
%	\frac{m\gd}{\kd} \simeq -\cos^2\!\theta \:T_1.
	m\gd / \kd \simeq -\cos^2\!\theta \:T_1.
\end{equation}
By fitting the experimentally-measured $\gd$ and $\kd$ to this formula near the charge transfer line, one can thus get a direct measure of the DQD $T_1$ time.  This shows an advantage of this technique over conventional charge-sensing:  as one is measuring dynamic phenomena (as opposed to simply the average value of the charge in the two dots), it is possible to directly extract important DQD timescales.

{\it Conclusions--} Using a somewhat novel master equation approach in conjunction with linear response, we have studied theoretically how charge dynamics in a DQD can cause damping and frequency shifts of a low-frequency mechanical resonator (such as an AFM tip).  Qualitatively new effects arise compared to the case of a single dot due to the cantilever's sensitivity to charge distribution, and due to the presence of coherent interdot tunneling.  We demonstrated that these effects could be used to detect and measure the magnitude of coherent tunneling, as well as extract the DQD $T_1$ time near the charge transfer line.  

We thank Y. Miyahara and L. Cockins for useful conversations, and acknowledge research support from CIFAR, NSERC and the McGill Centre for the Physics of Materials.

%%%%%%%%%%%%%%%%%%%%%%%%%%%%%%%%%%%%%%%%%%%%%%%%%%%%
%%%%%%%%%%%%%%%%%%%%%%%%%%%%%%%%%%%%%%%%%%%%%%%%%%%%
%%%%%%%%%%%%%%%%%%%%%%%%%%%%%%%%%%%%%%%%%%%%%%%%%%%%

\begin{appendix}

\section{Capacitance and coupling modelling}

We briefly describe in this appendix the modelling of parameters used to generate Fig.~1 and 2; we stress that the main results and equations of the paper are independent of this modelling.  To generate plots of AFM damping and frequency shift as a function of tip position $\vec{r}_{\rm tip}$, one needs to understand the dependence of the dimensionless gate voltages $\NN_\beta= -V_B C_{ \rm tip,\beta} ( | \vec{r}_{\rm tip} - \vec{r}_\beta|) /e$ on $\vec{r}_{\rm tip}$.  This dependence determines both the addition energies (~$E_{+\beta} = E_{\rm C \beta} \bigl ( 1-2 \mathcal{N}_{\beta} \bigr) -  E_{\rm C m} \mathcal{N}_{\bar{\beta}}$~) and the coupling strengths ($A_\beta=\partial E_{+ \beta}/\partial \z$).  We describe the dependence of $C_{\rm tip}$ on tip position using a simple functional form (see Eq.~(\ref{eq:alpha}) below) derived from the experimental results of Ref.~\onlinecite{Cockins10}; we also take parameter values in this form that correspond to typical values in those experiments.  
%We stress that this model is only used to calculate the simulated AFM images that are the insets of Fig. 1 and 2.

Following standard convention, we define the ``lever arm" $\alpha_\beta$ between the 2DEG and dot $\beta$ as:
\begin{equation}
 \alpha_\beta = \frac{C_{\rm tip,\beta}}{C_{\rm \Sigma,\beta}},
\end{equation}
where $C_{\Sigma,\beta}=e^2/(2 E_{\rm C \beta})$ is the total capacitance of the dot.  In general, $C_{\Sigma,\beta} \gg C_{\rm tip,\beta}$ and thus one can safely neglect the position dependence of $C_{\Sigma,  \beta}$.  Recent work reported in Refs.~\onlinecite{Cockins10} and \onlinecite{CockinsThesis} has addressed the variation of $\alpha$ with tip postion both experimentally and using finite difference modeling; the results can be approximated with a simple analytic form.  Considering a single, isolated dot and defining $\rho$ and $h$ such that $(\vec{r}_{\rm tip} - \vec{r}) = \vec{\rho} + h\hat{z}$, we approximate the lever arm as:
\begin{equation}
 \alpha(\rho,h) \simeq \frac{\alpha_0}{1 + \sqrt{(h/d_h)^2 +  (\rho/d_\rho)^2}}. \label{eq:alpha}
\end{equation}
Representative parameters are $\alpha_0 \sim 0.1$, $d_h\sim 10$~nm, and $d_\rho\sim 20$~nm.  Using this simple form and assuming moderate dot-specific parameter variation, we model the addition energies and coupling strengths for both dots at all oscillator positions.  \newline\indent

\section{Effective $T_1$ times due to electron tunnelling}

The effective TLS relaxation rate $1/T_1$ given in Eq.~(\ref{eq:tlsd}) describes relaxation due to electron tunnelling in the simplest case of spinless electrons, and where the 2DEG is symmetrically coupled to the two dots of the DQD.  Relaxing both these assumptions in our master equation treatment results in a modified form for the first (tunneling-induced) term in Eq.~(\ref{eq:tlsd}):
\begin{equation}
	T_1^{-1} \Big|_{\rm tunnel} = \bar{\Gamma}\left(1 - \cos^2\theta\: \Gamma_\delta^2 \right) \cosh(\beta\Delta)\Bigl( e^{-\beta|\bar{\epsilon}|} + 2e^{\beta(|\bar{\epsilon}|-E_m)}\Bigr)\label{eq:t1spinasym}.
\end{equation}
Here, tunnelling between the left (right) dot to the 2DEG is described by the total rate $\Gamma_L$ ($\Gamma_R$), and  we have defined 
$\bar{\Gamma} = (\Gamma_L +\Gamma_R)/2$, $\Gamma_\delta = (\Gamma_L -\Gamma_R)/(\Gamma_L +\Gamma_R)$.  The factor of two in the last term of the above expression reflects the presence of spin degeneracy. 

The above result (as well as Eq.~(\ref{eq:t1})) neglects the possibility of coherence in the 2DEG - DQD tunnelling (i.e.~the possibility of interference between tunnelling into a given DQD eigenstate via the left dot or via the right dot).  In the limit where the interdot spacing is much larger than the Fermi wavelength of the 2DEG, such interference terms are strongly suppressed.  In contrast, for a small interdot spacing, this interference will contribute.  One thus finds a different expression for the 2DEG-tunneling.  Taking $\Gamma_\delta=0$, the tunnelling contribution to the effective TLS relaxation rate $1/T_1$  becomes:
\begin{equation}
	T_1^{-1}\Big|_{\rm tunnel} = \Gamma \cosh(\beta\Delta)\Bigl(\cos^2\theta e^{-\beta|\bar{\epsilon}|} + (1+\cos^2\theta )e^{\beta(|\bar{\epsilon}|-E_m)}\Bigr) \label{eq:t1spinint},
\end{equation}
where the factors of $\cos\theta$ (cf.~Eq.~(\ref{eq:theta}) in the text) arise from electron interference.  This result suggests that the lead-mediated $T_1$ time is particularly long near the charge transfer line ($\cos\theta \sim 0$) due to destructive interference.  Because $T_{\rm 1,int}$ is likely to dominate the relaxation rate in this region, such an effect will have minimal influence on TLS damping in the one-electron DQD.  Nevertheless, this interference effect is closely intertwined with coherent tunneling in the DQD, and could offer an interesting topic for future investigations.  

\end{appendix}

%%%%%%%%%%%%%%%%%%%%%%%%%%%%%%%%%%%%%%%%%%%%%%%%%%%%
%%%%%%%%%%%%%%%%%%%%%%%%%%%%%%%%%%%%%%%%%%%%%%%%%%%%
%%%%%%%%%%%%%%%%%%%%%%%%%%%%%%%%%%%%%%%%%%%%%%%%%%%%

%\bibliography{ACTotalRefsV2}
\bibliographystyle{apsrev}
\bibliography{References_JGFinal}
%\bibliography{References_JGFinaletal}

\end{document}